# PHz-wide spectral interference through coherent plasma-induced fission of higher-order solitons

F. Köttig[1*], F. Tani[1], J. C. Travers[1,2] and P. St.J. Russell[1]

[1]Max Planck Institute for the Science of Light, Staudtstrasse 2, 91058 Erlangen, Germany
[2]School of Engineering and Physical Sciences, Heriot-Watt University, Edinburgh EH14 4AS, United Kingdom

We identify a novel regime of soliton-plasma interactions in which high-intensity ultrashort pulses of intermediate soliton order undergo coherent plasma-induced fission. Experimental results obtained in gas-filled hollow-core photonic crystal fibers are supported by rigorous numerical simulations. The cumulative blueshift of higher-order input solitons with ionizing intensities results in pulse splitting before the ultimate self-compression point, leading to the generation of robust pulse pairs with PHz bandwidths. The novel dynamics closes the gap between plasma-induced adiabatic soliton compression and modulational instability.

The interaction of ultrashort laser pulses with photo-induced plasmas has been the subject of intense research, and its understanding is of fundamental importance in many different fields, for example optical filamentation [1] and attosecond physics [2]. The field has been mostly studied in free space [3] and fiber capillaries [4]. In both these cases, however, spurious spatial effects complicate the dynamics and control over the propagation of the laser pulses is limited. Over the past few years, gas-filled hollow-core photonic crystal fiber [5] (PCF) has matured as an ideal platform for studying the interaction of laser light with matter. The tight confinement of high-intensity femtosecond pulses in a single spatial mode with precisely controllable anomalous dispersion allows established soliton dynamics to be extended into the strong-field regime [6], opening up unique possibilities. For the first time, it has been possible to study how optical solitons interact with plasmas over long path-lengths and broad frequency ranges in a well-controlled environment, as well as effects such as plasma-induced blue-shifting [7,8], adiabatic soliton compression [9] and modulational instability (MI) [10].

In previous work, soliton-plasma interactions were mainly studied in the regimes of low and very high soliton order. Here we investigate high-intensity ultrashort solitons of intermediate order, and demonstrate novel coherent plasma-induced fission, leading to pulse splitting and the production of robust pulse pairs of PHz bandwidth that co-propagate phase-locked over cm-long distances. We furthermore show that the regimes of soliton-based pulse compression [11–13], supercontinuum generation [14,15], dispersive wave (DW) emission [16,17], soliton blue-shifting and plasma-induced fission can be accessed in a single system, simply by changing the input pulse energy. The findings are of great practical interest as hollow-core PCF-based pulse manipulation schemes evolve into a highly-demanded asset in modern high-repetition rate laser systems [18,19].

In the experiment, an 8-cm-long kagomé-type PCF with 36 μm core diameter and 200 nm core wall thickness was placed in a gas-cell filled with 2 bar of Kr. It was pumped by 28-fs-long (full-width-half-maximum - FWHM) pulses at ~0.29 PHz (1.03 μm) with up to 13.2 μJ pulse energy at a repetition rate of 151 kHz. The fiber has a loss of ~1 dB/m at the pump frequency, and through careful alignment of the incoupling, only the fundamental $LP_{01}$-like mode was excited. As the complex temporal evolution of the pulses and the plasma formation along the fiber are not directly accessible in the experiment, we have to rely on simulations to understand them in detail. We therefore numerically modeled the pulse propagation using a single-mode unidirectional field equation [20], including photoionization with the Perelomov, Popov, Terent'ev (PPT) ionization rates [21], modified with the Ammosov, Delone, Krainov (ADK) coefficients [22]. The dispersion of the fiber was modelled using a simple capillary model [6,23], modified by a wavelength-dependent effective core radius that allows an accurate estimate of the modal refractive index at longer wavelength [24]. The $s$-parameter in this modified capillary model was set to $s = 0.08$, an optimum value that was determined by finite-element modelling of an idealized fiber with the same structural parameters. Since the input pulses were measured by frequency-resolved optical gating (FROG), there were no free parameters in the simulations.

When the fiber is filled with 2 bar of Kr, the zero-dispersion point (ZDP) is located at 0.57 PHz, so that the pump pulses lie in the anomalous dispersion region and can therefore form solitons. Their short duration means that the pulses can have high energies while keeping the soliton order $N$ well below the MI regime [25] ($N < 7.1$ in the experiment). Due to the low soliton order, and because the MI gain bandwidth quickly falls within the expanding spectrum of the pump pulses, the dynamics are dominated by coherent soliton fission rather than incoherent MI. Figure 1(a) shows the experimental spectra at the fiber output. The following phenomena can be identified: soliton self-compression [12,13] (for input energies smaller than 4 µJ), supercontinuum generation [15], DW emission [16,17] and soliton blue-shifting [8] (between 4 and 8 µJ), and finally a novel effect: plasma-induced soliton fission for input energies greater than 8 µJ. These effects will be explained in the following, and the origin of the plasma-induced soliton fission identified.

With increasing input pulse energy, the combined effects of self-phase modulation (SPM) and anomalous dispersion result in soliton self-compression [26] to a duration $\propto 1/N$ [27], the concomitant spectral broadening resulting in a multi-octave-spanning supercontinuum. The optical shock effect at the trailing edge of the pulses strongly enhances the high-frequency side of the spectrum, resulting in efficient resonant transfer of energy to a phase-matched DW [28] in the ultraviolet (UV). According to the simulations, the intensity exceeds $10^{14}$ W/cm$^2$ at the point of maximum temporal compression, where plasma densities greater than $10^{18}$ cm$^{-3}$ are generated inside the fiber. The loss associated with the generation of free electrons results in a drop in transmission (Fig. 1(b)) by more than 30% for input energies greater than 4 µJ. Photoionization of the Kr gas inside the fiber leads to the emission of a blue-shifted soliton [8,29], extending from ~0.35 PHz out to more than 0.5 PHz. This blue-shifted soliton emits a second DW, and as it shifts towards the ZDP with increasing input energy, so does the DW because of phase matching.

While the dynamics of this process are well established, they change considerably for input energies greater than 8 µJ when, as opposed to blue-shifting of single solitons, well-defined spectral interference fringes are observed over an extremely broad frequency range, extending into the vacuum UV. These fringes suggest the presence of a pulse *pair* in the time domain (if there were more than two equally spaced pulses in the pulse train, the

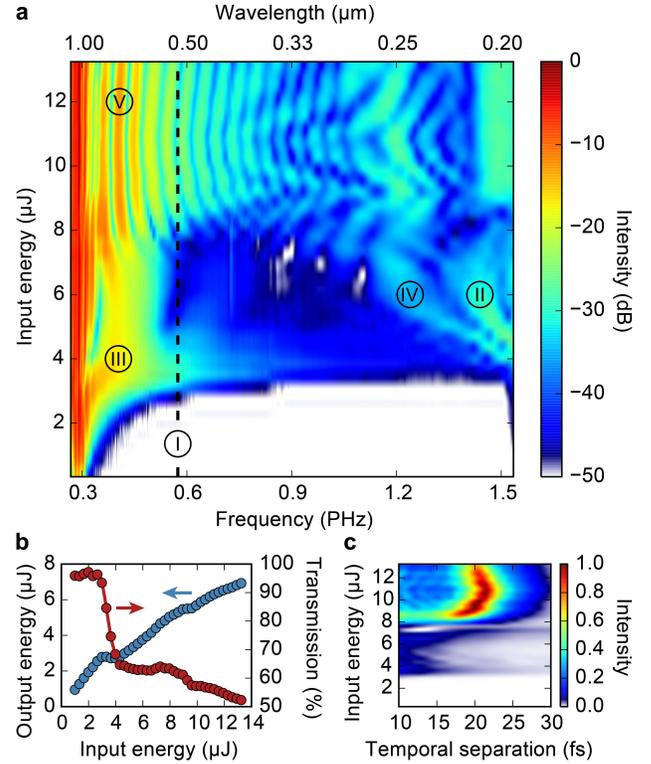

FIG. 1 (a) Experimental output spectra, measured over 5 ms integration time (>750 laser shots). Using a conventional Si CCD spectrometer, only the spectral range from 0.27 to 1.54 PHz (1.12 to 0.2 µm) could be measured. Dispersion is anomalous on the low-frequency side of the ZDP (I). A DW is emitted at a phase-matched frequency in the UV (II). Ionization of the filling gas leads to emission of a blue-shifted soliton (III), which generates a second DW at lower frequencies (IV). At high input energies, distinct interference fringes are visible across the entire spectrum (V). (b) Output energy and transmission of the fiber (including in-coupling losses). (c) Temporal separation of the pulses in a pulse pair corresponding to the spectrum in (a), retrieved via Fourier transform for frequencies above 0.39 PHz. When distinct interference fringes occur in the output spectra (region (V) in (a)), they have a spectral spacing of ~50 THz, which corresponds to a temporal separation of ~20 fs.

interference fringes would exhibit weaker local maxima between the peaks), with a temporal separation inversely proportional to the spectral fringe spacing, as shown in Fig. 1(c). We measured the output pulses of the fiber using FROG (Fig. 2). For moderate input energies, the pulses strongly self-compress in the fiber, but remain as one unit (Fig. 2(a)). For higher input energies, however, pulse pairs emerge (Fig. 2(b),(c)) with a temporal separation of 20 fs, in agreement with the spacing (50 THz) of the interference fringes in the spectrum (Fig. 1). Even though the FROG measurement took around 15 minutes (at 1 fs temporal resolution), the example trace shown in Fig. 2(d) does not exhibit any fluctuations, confirming that the underlying soliton-plasma dynamics are both coherent and stable.

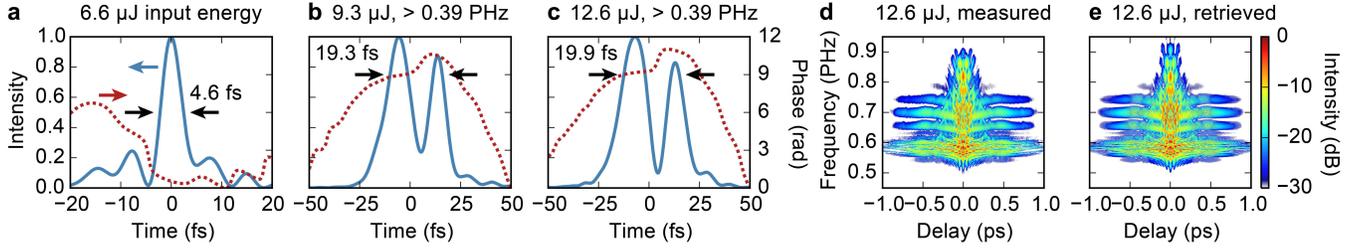

FIG. 2 Pulse measurement at the output of the fiber via noncollinear second harmonic generation (SHG) FROG in a 10-μm-thick beta-barium-borate (BBO) crystal. The dispersion in the measurement path was numerically corrected after retrieval of the pulses. Despite the thin crystal, only part of the multi-octave-spanning spectra could be measured, but the conclusions drawn from these measurements are representative of a wider spectral range, given that they agree with the independent measurements shown in Fig. 1. (a) Temporal pulse-shape before the onset of pulse splitting, with a FWHM duration of 4.6 fs, limited by the bandwidth of the FROG device. (b),(c) Temporal pulse-shape in the input energy regime where pulse splitting occurs, Fourier filtered above 0.39 PHz to suppress the residual pump pulse, which did not compress strongly. (d),(e) Measured and retrieved FROG traces corresponding to (c). The FROG trace was retrieved on a 2048 x 2048 grid, with an error of $G = 0.95\%$.

Figure 3 shows the simulated pulse evolution and ionization rate at input energies representative of the different regimes observed in the experiment, i.e., for moderate (5 μJ) and high (13 μJ) input energy. At moderate energies the input pulses undergo clean self-compression (Fig. 3(a)), reaching durations as short as 1.2 fs (FWHM), with less than one optical cycle under the intensity envelope. Under these circumstances, significant ionization occurs only close to the point of maximum temporal compression (Fig. 3(c)), where the free electron population effectively builds up during a single cycle of the electric field (at this point, the Keldysh parameter [30] $\gamma < 0.7$, i.e., ionization occurs in the nonadiabatic tunneling regime). The resulting ionization-induced frequency blueshift $\Delta\omega$ of the pulses is given by [31]:

$$\frac{\partial \Delta\omega(z,t)}{\partial z} = \frac{\omega_0}{2n_0 c \rho_{cr}} \frac{\partial \rho(z,t)}{\partial t}, \quad (1)$$

where $z$ is propagation distance, $t$ the time, $\omega_0$ the central frequency of the pulse, $n_0$ the linear refractive index at $\omega_0$, $c$ the speed of light in vacuum, $\rho$ the plasma density and $\rho_{cr}$ the critical density at which the plasma becomes opaque at $\omega_0$. Since the plasma frequency blueshift is proportional to the ionization rate $\partial \rho/\partial t$, the rapid plasma build-up across the pulse (Fig. 3(c), where the ionization rate reaches $10^{19}$ cm$^{-3}$ fs$^{-1}$) causes strong blue-shifting around the maximum temporal compression point. In the time-domain this leads to an acceleration of the pulse in the anomalous dispersion region (Fig. 3(a), where the pulse curves to the left after the point of maximum compression). This perturbation is strong enough to emit a blue-shifted soliton (Fig. 1(a)).

When the input energy is high enough, however, self-compression of the input pulses is not needed for significant ionization to occur. It can even occur at the fiber input (Fig. 3(d)), so that plasma densities greater than $10^{18}$ cm$^{-3}$ can be maintained over distances of several cm. During the initial stage of propagation, the ionization builds up progressively over several cycles of the electric field. As a result the plasma density grows slowly, so that the perturbation induced by the blueshift is too weak to eject a soliton.

Instead, the pulse experiences asymmetric spectral broadening [10], caused by the combined effects of the optical Kerr nonlinearity and plasma-induced phase-modulation (Fig. 4(a)). In contrast to conventional SPM, high frequencies are generated at the leading edge of the pulse and then accelerate away from it due to their higher group velocity in the anomalous dispersion region. Traditional soliton self-compression is suppressed at this point. As recombination of free electrons occurs on time-scales much longer than the pulse duration, the plasma generated at the leading edge of the pulse also acts on its trailing edge by reducing its blueshift and slowing down its acceleration [32]. This leads to a temporal separation, which increases with propagation distance and eventually splits the pulse into sub-pulses (Fig. 4(b)). For the 13 μJ pulse shown in Fig. 3(b), two sub-pulses emerge, separated by 14 fs at their point of generation. These sub-pulses still have sufficient energy to individually undergo soliton self-compression, culminating in two almost identical single-cycle pulses with PHz bandwidth, giving rise to spectral interference fringes. Note that the observed dynamics differ from the plasma-induced MI effects theoretically investigated by Saleh et al. [10], where a shower of individual solitons is generated through MI. Since the pump pulses in our work are sub-30-fs in duration, high intensities can be reached at moderate soliton order, resulting in fully coherent dynamics and a well-defined pulse splitting.

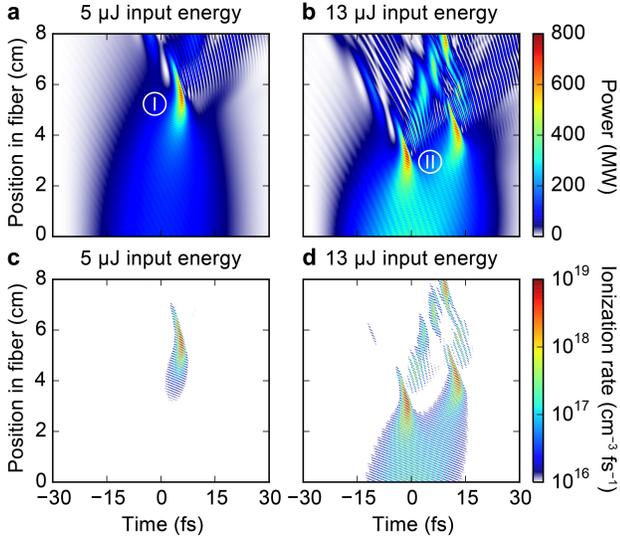

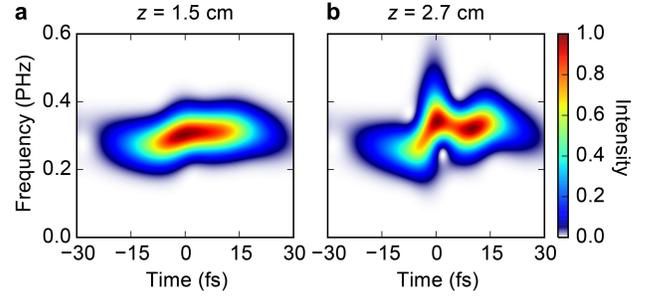

FIG. 4 Time-frequency analysis (spectrogram) of the 13 μJ pulse shown in Fig. 3(b), at $z = 1.5$ cm and $z = 2.7$ cm along the fiber, calculated using a 5-fs-long Gaussian gate-pulse. (a) The pulse experiences asymmetric spectral broadening towards high frequencies due to ionization at its leading edge. (b) Upon further propagation, the pulse breaks up and two sub-pulses emerge.

FIG. 3 Simulated pulse dynamics for moderate (5 μJ) and high (13 μJ) input energy. The input pulse was measured via FROG. (a),(b) Evolution of the temporal pulse-shape with propagation distance. At moderate input energy, the input pulse undergoes clean self-compression (I), whereas at high input energy, it splits into two sub-pulses (II), which then individually undergo self-compression. (c),(d) Evolution of the ionization rate $\partial \rho(z,t)/\partial t$. It closely follows the temporal pulse-shape, but significant values only occur when the pulse is sufficiently intense (>300 MW of power).

Although the simulations reveal the mechanism of the experimentally observed pulse splitting, the generation of pulse pairs appears to be more robust in the experiment, where consistent dynamics are observed over a wide range of input energies. We attribute this to the high pulse repetition rate (151 kHz), which leads to dynamics on a longer timescale, potentially caused by slow thermal gas dynamics. The associated dynamics are complicated and will require further research. In the experiment, the regimes of soliton blue-shifting and plasma-induced fission can be clearly distinguished. Also, the transmission of the fiber exhibits a slight revival at an input energy of ~8 μJ, coinciding with the appearance of spectral fringes. Simulations show that this is caused by significant ionization at an early stage along the fiber. As a consequence self-compression is strongly disrupted, resulting in lower overall ionization, higher transmission, and even intermittent suppression of DW emission. At even higher energies, although the pulses break up before they reach the point of maximum compression, the emerging sub-pulses still have sufficient energy to individually self-compress to the single-cycle regime while maintaining a fixed relative phase. As pulse break-up and subsequent self-compression occurs closer to the input of the fiber with increasing input energy, the generated pulse pairs propagate over an increased length as well. Despite an increase in input energy of more than 60% (from 8 μJ to more than 13 μJ), the spectral interference fringes persist, and the temporal shape, phase and spacing of the individual pulses in the pulse pairs remains constant (Fig. 2(b),(c)).

Further experiments with different fiber lengths showed very consistent dynamics, even for a fiber that was 2 cm shorter, confirming that the pulse pairs generated upon plasma-induced soliton fission propagate phase-locked over long distances. These effects are not unique to Kr gas, but can be reproduced experimentally in the lighter noble gases Ar and Ne, albeit at higher energies since they have higher ionization potentials.

In conclusion, a novel regime of high-intensity soliton-plasma dynamics results in plasma-induced fission of higher-order solitons, generating robust pairs of ultrashort pulses with PHz bandwidth. The interaction of free electrons within an ultrashort pulse causes coherent and well-defined pulse break-up, far from the regime of MI and for soliton orders not greater than 7. These novel results also delineate the parameter range where clean soliton self-compression occurs, i.e., where efficient pulse compression can be achieved in gas-filled hollow-core PCF. Moreover, the system could be used as a pulse shaper to create a precise replica of an ultra-broadband pulse with a well-defined delay.

[1] A. Couairon and A. Mysyrowicz, Phys. Rep. **441**, 47–189 (2007).
[2] F. Krausz and M. Ivanov, Rev. Mod. Phys. **81**, 163–234 (2009).
[3] L. Bergé, S. Skupin, R. Nuter, J. Kasparian, and J.-P. Wolf, Rep. Prog. Phys. **70**, 1633–1713 (2007).
[4] A. Rundquist, C. G. Durfee, Z. Chang, C. Herne, S. Backus, M. M. Murnane, and H. C. Kapteyn, Science **280**, 1412–1415 (1998).
[5] P. St.J. Russell, J. Lightwave Technol. **24**, 4729–4749 (2006).
[6] P. St.J. Russell, P. Hölzer, W. Chang, A. Abdolvand, and J. C. Travers, Nat. Photon. **8**, 278–286 (2014).
[7] A. B. Fedotov, E. E. Serebryannikov, and A. M. Zheltikov, Phys. Rev. A **76**, 053811 (2007).
[8] P. Hölzer, W. Chang, J. C. Travers, A. Nazarkin, J. Nold, N. Y. Joly, M. F. Saleh, F. Biancalana, and P. St.J. Russell, Phys. Rev. Lett. **107**, 203901 (2011).
[9] W. Chang, P. Hölzer, J. C. Travers, and P. St.J. Russell, Opt. Lett. **38**, 2984–2987 (2013).
[10] M. F. Saleh, W. Chang, J. C. Travers, P. St.J. Russell, and F. Biancalana, Phys. Rev. Lett. **109**, 113902 (2012).
[11] D. G. Ouzounov, C. J. Hensley, A. L. Gaeta, N. Venkataraman, M. T. Gallagher, and K. W. Koch, Opt. Express **13**, 6153–6159 (2005).
[12] K. F. Mak, J. C. Travers, N. Y. Joly, A. Abdolvand, and P. St.J. Russell, Opt. Lett. **38**, 3592–3595 (2013).
[13] T. Balciunas, C. Fourcade-Dutin, G. Fan, T. Witting, A. A. Voronin, A. M. Zheltikov, F. Gerome, G. G. Paulus, A. Baltuska, and F. Benabid, Nat. Commun. **6**, 6117 (2015).
[14] A. V. Husakou and J. Herrmann, Phys. Rev. Lett. **87**, 203901 (2001).
[15] A. Ermolov, K. F. Mak, M. H. Frosz, J. C. Travers, and P. St.J. Russell, Phys. Rev. A **92**, 033821 (2015).
[16] N. Y. Joly, J. Nold, W. Chang, P. Hölzer, A. Nazarkin, G. K. L. Wong, F. Biancalana, and P. St.J. Russell, Phys. Rev. Lett. **106**, 203901 (2011).
[17] K. F. Mak, J. C. Travers, P. Hölzer, N. Y. Joly, and P. St.J. Russell, Opt. Express **21**, 10942–10953 (2013).
[18] F. Emaury, A. Diebold, C. J. Saraceno, and U. Keller, Optica **2**, 980–984 (2015).
[19] M. Gebhardt, C. Gaida, S. Hädrich, F. Stutzki, C. Jauregui, J. Limpert, and A. Tünnermann, Opt. Lett. **40**, 2770–2773 (2015).
[20] F. Tani, J. C. Travers, and P. St.J. Russell, J. Opt. Soc. Am. B **31**, 311–320 (2014).
[21] A. M. Perelomov, V. S. Popov, and M. V. Terent'ev, Sov. Phys. JETP **23**, 924 (1966).
[22] F. A. Ilkov, J. E. Decker, and S. L. Chin, J. Phys. B **25**, 4005 (1992).
[23] E. A. J. Marcatili and R. A. Schmeltzer, Bell Syst. Tech. J. **43**, 1783–1809 (1964).
[24] M. A. Finger, N. Y. Joly, T. Weiss, and P. St.J. Russell, Opt. Lett. 39, 821–824 (2014).
[25] J. M. Dudley, G. Genty, and S. Coen, Rev. Mod. Phys. **78**, 1135–1184 (2006).
[26] L. F. Mollenauer, R. H. Stolen, J. P. Gordon, and W. J. Tomlinson, Opt. Lett. **8**, 289–291 (1983).
[27] C.-M. Chen and P. L. Kelley, J. Opt. Soc. Am. B **19**, 1961–1967 (2002).
[28] N. Akhmediev and M. Karlsson, Phys. Rev. A **51**, 2602 (1995).
[29] M. F. Saleh, W. Chang, P. Hölzer, A. Nazarkin, J. C. Travers, N. Y. Joly, P. St.J. Russell, and F. Biancalana, Phys. Rev. Lett. **107**, 203902 (2011).
[30] L. V. Keldysh, Sov. Phys. JETP **20**, 1307–1314 (1965).
[31] P. B. Corkum, IEEE J. Quant. Electron. **21**, 216–232 (1985).
[32] M. F. Saleh and F. Biancalana, Phys. Rev. A **84**, 063838 (2011).